# The Role of Agricultural Sector Productivity in Economic Growth: The Case of Iran's Economic Development Plan

Morteza Tahamipour[1] & Mina Mahmoudi[2,*]

[1]Department of Agricultural Economics, University of Beheshti, Tehran, Iran

[2]Department of Economics, University of Nevada, Reno

*Corresponding author: Department of Economics, University of Nevada, Reno. E-mail: mmahmoudi@ur.edu

**Abstract**

This study provides the theoretical framework and empirical model for productivity growth evaluations in agricultural sector as one of the most important sectors in Iran's economic development plan. We use the Solow residual model to measure the productivity growth share in the value-added growth of the agricultural sector. Our time series data includes value-added per worker, employment, and capital in this sector. The results show that the average total factor productivity growth rate in the agricultural sector is -0.72% during 1991-2010. Also, during this period, the share of total factor productivity growth in the value-added growth is -19.6%, while it has been forecasted to be 33.8% in the fourth development plan. Considering the effective role of capital in the agricultural low productivity, we suggest applying productivity management plans (especially in regards of capital productivity) to achieve future growth goals.

**Keywords:** Productivity, Agricultural Sector, Development, Iran

## 1. Introduction

In our modern world, considering productivity improvement by the governments is not a choice but a necessity due to limited resources, unlimited human needs, increasing population and intensive market competition. Without any doubt, growth in developing countries depends mainly on productivity and increasing the productivity growth rate. Measuring and analyzing productivity indexes would help us to recognize shortcomings and improvement opportunities for the goal of higher production and a long-run economic growth.

According to the production and supply theories, production growth in a sector would be possible in two ways: the use of more production factors and the use of more advanced technologies. In Iran and several other developing countries, the limitation of water and other agricultural inputs restricts the first way of increasing production in the long-run (Amini et. al, 2016; Karami et. al, 2016). Therefore, considering the second way of production growth (increasing factor productivity) is a necessity for increasing the supply of products (Amini, 2005; Amini & Neshat, 2007).

The five-year development plans in Iran focus on the interactions of private and public sectors, their investment requirements, structural reforms, and stabilization policies to accelerate economic, social and cultural development (Mahmoudi & Chizari, 2013). The fourth economic development plan in Iran (2006-2010) has determined quantitative goals to increase the total factor productivity as a way to achieve sustainable economic growth for the first time.

One of the most important economic sectors in Iran is the agricultural sector considering its share in the gross domestic product (Karami & Mahmoudi, 2013). In the fourth development plan, the average growth of the agricultural sector value-added was considered to be 6.5% annually, where 4.3% has been predicted to be attained by increase in the level of investment and 2.2% (equal to 33.8% of the value-added growth) to be attained by the total factor productivity growth. In the other words, in the fourth development plan, it has been predicted to obtain one third of economic growth by increasing the total factor productivity. Also, the partial productivity growth rate of labor and capital in the agricultural sector, during the years of the fourth development plan, considered to be respectively 4.6% and 0.1% (Management and Planning Organization of Iran, 2004).

Focusing on the importance of having a strategic plan to improve productivity growth in different economic sectors, this study provides the theoretical framework and empirical model for productivity growth evaluation aims to achieve the productivity goals in the agricultural sector as one of the most important economic sectors in Iran's development plan. The rest of the paper is organized as follows. Section 2 describes the data and methodology of the study. Section 3 reports the empirical results. A summary and conclusion is provided in section 4.

## 2. Methodology and Data

There have been several methods presented in the last decades for measuring the total factor productivity. However, there are two general methods suggested in economics literature including parametric and non-parametric methods. In the parametric method, productivity would be calculated by estimating production, cost or profit functions. In the non-parametric method, productivity would be determined using mathematical programming, calculating index numbers or by growth accounting methods (Diewert, 1981 & 1992). In this study, due to the goal of assessing agricultural sector productivity focusing on the period of the fourth development plan, and also considering the limitation of the data statistics, we used the growth accounting method which has been suggested by the Asian Productivity Organization (Asian Productivity Organization, 2004).

Based on the growth accounting method, we can consider the output growth derived from an input growth (like labor or capital) and a change in the total productivity. Standard growth theory assumes that the economy's level of production depends on the economy's employment level, capital level, and level of technology (Mahmoudi, 2017). Let

$$(1) \quad Q_t = A_t F(K_t, L_t)$$

be the production function for the model economy, where $Q_t$ shows the output level, $L_t$ and $K_t$ respectively show the labor and capital stock level at time *t*, and $A_t$ shows the term for technology development or total factor productivity.

There are some important assumptions in the growth accounting method. The first is that the term of technology development or total factor productivity ($A_t$), as we show in equation (1), is separable. The second assumption is that the production function has a constant return to scale. The third is that the market is in perfect competition and the producers are price takers. Finally, the fourth assumption is that the producers' goal is to maximize their profit.

If we differentiate the production function in equation (1) with respect to time, we obtain

$$(2) \quad \frac{dQ}{dt} = \frac{dA}{dt} F(K_t, L_t) + A_t \frac{\partial F}{\partial K} \frac{dK}{dt} + A_t \frac{\partial F}{\partial L} \frac{dL}{dt}$$

By dividing both sides on $Q_t$, we get

$$(3) \quad \frac{dQ}{dt}/Q_t = \frac{dA}{dt}/A_t + \frac{\partial F}{\partial K}\frac{dK}{dt}/F(K_t, L_t) + \frac{\partial F}{\partial L}\frac{dL}{dt}/F(K_t, L_t) .$$

If we show the output elasticity with respect to the labor and capital production factors respectively by $W_L$ and $W_K$, then

$$(4)\ W_L = \frac{\partial Q}{\partial L}\frac{L}{Q} = A_t\frac{\partial F}{\partial L}\frac{L}{Q}$$

$$(5)\ W_K = \frac{\partial Q}{\partial K}\frac{K}{Q} = A_t\frac{\partial F}{\partial K}\frac{K}{Q}$$

By replacing the elasticity equations in equation (3), we obtain

$$(6)\ \frac{dQ}{dt}/Q_t = \frac{dA}{dt}/A_t + W_K\frac{dK}{dt}/K + W_L\frac{dL}{dt}/L$$

We can rewrite equation (6) as follows where dot signs show the growth rate of different variables:

$$(7)\ \frac{\dot{Q}}{Q} = \frac{\dot{A}}{A} + W_K\frac{\dot{K}}{K} + W_L\frac{\dot{L}}{L}$$

In equation (7), $\frac{\dot{A}}{A}$ presents the total factor productivity growth rate and as it is presented in the following equation, we can calculate it as a residual:

$$(8)\ \frac{\dot{A}}{A} = \frac{\dot{Q}}{Q} - W_K\frac{\dot{K}}{K} - W_L\frac{\dot{L}}{L}$$

The output productivity obtained using this method, shows a portion of production growth that cannot be presented by labor and capital changes, but it is related to the changes in total factor productivity growth (Mawson et. al, 2003). In the productivity literature, this model is called "the Solow Residual Model" and can be simply rewritten as

$$(9)\ \widehat{TFP} = \hat{Q} - \alpha\hat{K} - \beta\hat{L}$$

where Q, K, and L respectively show the output value-added, capital and employment growth rates. The capital and labor shares in the production are respectively shown by $\alpha$ and $\beta$. If we apply the assumption of constant returns to scale, then $\beta = 1 - \alpha$ (Solow, 1988).

The Solow Residual Model is simply the difference between the weighted average of the factors' growth and the production growth. In the other words, the total factor productivity is considered as the aggregate productivity changes in capital and labor (Salami, 1997). This latter sentence can be mathematically shown as

$$(10)\ \widehat{TFP} = \alpha\widehat{APK} - \beta\widehat{APL}$$

where $\widehat{APK}$ is the aggregate productivity of capital and $\widehat{APL}$ is the aggregate productivity of labor. If we replace equation (9) in (10), considering that the production elasticities of labor and capital are respectively *α* and *β*, we obtain

$$(11) \quad \hat{Q} = \alpha\hat{K} + \beta\hat{L} + \alpha\widehat{APK} + \beta\widehat{APL}$$

This equation divides the production growth into two parts: the first two terms show the share of changes in capital and labor in the production growth, and the other two terms are related to the share of total factor productivity in the production or value-added growth.

In this study, the agricultural sector is chosen for productivity estimation as a representative of the other economic sectors due to its considerable size. The productivity growth of this sector is evaluated considering the fourth development plan goals. The data that we use for this purpose includes the output value-added in 1997 billion Rials, the capital stock value in 1997 billion Rials, and the employment level (persons) during the period of 1991-2010. The related data for the agricultural sector is obtained from the Central Bank of Iran (Central Bank of Iran, 2010).

One of the several methods to calculate the capital and labor shares in the Solow Residual Model is using the input-output tables. In this method, the share of compensation of employees in the output value-added is considered to be a substitute for the share of labor in production by applying some adjustments(Note 1). In this study, using the information from the input-output tables (Iran Statistics Center, 2010) and considering the assumption of the constant returns to scale, the shares for the capital and labor have been estimated.

## 3. Results

Table 1 presents the results for the estimation of the total factor productivity growth rate in the agricultural sector together with the partial productivity for capital and labor in this sector. As it can be seen in this table, the agricultural sector productivity growth rate was negative in most of the years. This could be related to different shocks in the production process in these years which is important to investigate carefully. For example, the large fall in the agricultural sector productivity growth rate of the year 1999, is most probably related to the considerable drought happened in this year. The results show that average productivity growth rate in the agricultural sector during the years 1991-2010 is -0.72 percent. The average partial productivity growth rate of the capital and labor in the agricultural sector during these years are respectively -2.89 and 2.76 percent.

**Table 1.** Capital, Labor, and Total Factor Productivity Growth in the Agricultural Sector

| Year | Capital productivity growth rate (%) | Labor productivity growth rate (%) | Total factor productivity growth rate (%) |
|------|--------------------------------------|------------------------------------|-------------------------------------------|
| 1992 | 3.14   | 9.68  | 5.65   |
| 1993 | -5.78  | 0.36  | -3.42  |
| 1994 | -3.13  | 1.10  | -1.51  |
| 1995 | 0.10   | 2.11  | 0.87   |
| 1996 | -3.13  | 2.12  | -1.11  |
| 1997 | -4.91  | -0.08 | -3.05  |
| 1998 | 6.70   | 9.01  | 7.59   |
| 1999 | -15.55 | -6.91 | -12.23 |
| 2000 | -2.38  | 2.82  | -0.39  |
| 2001 | -8.14  | -1.59 | -5.63  |
| 2002 | 4.69   | 10.26 | 6.82   |
| 2003 | -2.18  | 3.98  | 0.19   |
| 2004 | -8.63  | -0.89 | -5.66  |
| 2005 | -0.92  | 6.88  | 2.08   |
| 2006 | -3.18  | 2.56  | -0.98  |
| 2007 | 1.41   | 6.89  | 3.31   |
| 2008 | -1.89  | 3.89  | 0.96   |
| 2009 | -4.68  | 0.73  | -2.07  |
| 2010 | -6.39  | -0.46 | -5.11  |

During the fourth economic development plan (2006-2010), the average productivity growth rate in the agricultural sector is -0.78, which is still a negative unacceptable value. During these years, the average partial productivity growth rate of capital and labor in the agricultural sector are respectively -2.95 and 2.72 percent. Due to the negative value for the capital average partial productivity growth rate during the fourth development plan, it is obvious that there should be a more responsible focus on boosting the capital productivity.

The share of the capital and labor factors in the production, considering the input-output table of the year 2010, are respectively 62 and 38 percent. Assuming the constancy of these shares during the period of our interest, we were able to calculate the share of total factor productivity growth in the production growth (how much total factor productivity growth helps the production growth) using the Solow Residual Model. First, we divided the output value-added growth into the capital and labor factors growth and their partial productivity growth. Then by considering the total output value-added growth rate to be 100, the share of each factor and partial productivities were calculated from the total output value-added growth. Table 2 shows the results of these calculations. According to the Solow Residual Model, the sum of capital and labor partial productivity growth shares is equal to the share of total factor productivity growth in the output value-added growth.

**Table 2.** Productivity and Input Growth Shares in the Agricultural Value-Added Growth

|  | Average growth rate (%) | Share (%) |
|---|---|---|
| Output value-added | 3.67 | 100 |
| Capital | 0.35 | 9.54 |
| Labor | 4.04 | 110.1 |
| Total factor productivity | -0.72 | -19.62 |

As it can be seen in Table 2, the total factor productivity has not helped the output value-added of the agriculture sector to grow during the years 1991-2010. The negative effect of the total factor productivity on the agriculture sector value-added growth is due to the negative partial productivity growth rate of capital in this sector. Therefore, future plans to improve the capital productivity can increase the share of the total factor productivity in the value-added growth.

**Table 3.** Productivity and Input Growth Shares in the Agricultural Value-Added Growth During the Fourth Economic Development Plan (2006-2010)

|  | Average growth rate (%) | Share (%) |
|---|---|---|
| Output value-added | 3.76 | 100 |
| Capital | 0.41 | 9.31 |
| Labor | 4.13 | 107.45 |
| Total factor productivity | -0.78 | -19.15 |

Table 3 shows the shares of total factor productivity and input growth in the agricultural value-added growth during the fourth economic development plan (2006-2010). Same as the whole period of the study, the total factor productivity growth during the fourth development plan has a negative effect on the agriculture sector value-added growth. Also, it can be seen that the share of capital growth in the output value-added growth is so small comparing to the share of labor growth.

## 4. Conclusion

In this study, we investigate the theoretical framework and empirical model to estimate the productivity growth rate in the agricultural sector as one of the most important sectors in Iran's economic development plan. The results show that during the years 1991-2010, the value-added growth in the agricultural sector had obtained mainly by the investment improvements especially through employment, while the productivity growth did not have a positive effect on the value-added growth. Therefore, productivity management plans should be applied in order to increase the overall productivity by improving the way physical and human capital is used in the agricultural sector and its subsectors.

Comparing the results of the study with the aim determined in the fourth development plan, we see that the total factor productivity growth had a negative share in the value-added growth during the period 2006-2010, while it was determined to contribute 33.8 percent of the value-added growth in the agricultural sector. This shows that the fourth development plan goal in regards of the agricultural sector has not been satisfied in the past. Therefore, in order to achieve the goal of productivity in the future development plans, the agricultural sector must correct its old policies in regards of quantitative changes in the investments and it should focus on improving the capital productivity.

The role of government and the size of that relative to the whole economy is another important fact to be considered in Iran as a developing country. "Without government providing the rule of law and protecting private property, productive behavior will not flourish privately", (Pingle & Mahmoudi, 2016). As a less developed country, Iran can most likely benefit from an increase in government size (to a specific point) and a more organized focus of government on the role of individual sectors in increasing productivity.

**Note**

Note 1. Because the share of the family labors in the mixed revenue is hidden and inseparable, 50% of the mixed revenue was considered as the family labor share and was added to the agricultural sector compensation of employees.